\documentclass[aps,twocolumn,groupedaddress,showpacs]{revtex4}

\begin{document}
% You should use BibTeX and apsrev.bst for references
%\bibliographystyle{apsrev}

%\preprint{}

\title{Quantum orders in an exact soluble model}
% Optional argument for running titles on pages
%\title[]{}


\author{Xiao-Gang Wen}
%\email{Your e-mail address}
\homepage{http://dao.mit.edu/~wen}
%\thanks{}
\affiliation{Department of Physics, Massachusetts Institute of Technology,
Cambridge, Massachusetts 02139}


%\date{\today}
\date{April 2002}

\begin{abstract}
We find all the exact eigenstates and eigenvalues of a spin-1/2 model on
square lattice: $ H=16g \sum_{\v i} S^y_{\v i} S^x_{\v i+\hat{\v x}} S^y_{\v
i+\hat{\v x}+\hat{\v y}} S^x_{\v i+\hat{\v y}}$.  We show that the ground
states for $g<0$ and $g>0$ have different quantum orders described by Z2A and
Z2B projective symmetry groups.  The phase transition at $g=0$ represents a
new kind of phase transitions that changes quantum orders but not symmetry.
Both the Z2A and Z2B states are described by $Z_2$ lattice gauge theories at
low energies. They have robust topologically degenerate ground states and
gapless edge excitations.
\end{abstract}

\maketitle

%\section{Introduction}
%\label{intro}

\noindent
{\bf Introduction}

We used to believe that all phases of matter are described by Landau's
symmetry breaking theory.\cite{L3726,GL5064} The symmetry and the related
order parameters dominate our understanding of phases and phases transitions
for over 50 years.  In this respect, the fractional quantum Hall (FQH) states
discovered in 1982\cite{TSG8259,L8395} opened a new chapter in condensed
matter physics. The theory of phases and phase transitions entered into a new
era. This is because all different FQH states have the same symmetry and hence
cannot be described by the Landau's theory.  In 1989, it was realized that FQH
states, having a robust topological degeneracy, contain a completely new kind
of order - topological order.\cite{Wtop} A whole new theory was developed to
describe the topological orders in FQH liquids. (For a review, see
\Ref{Wtoprev}.) 

The Landau's theory was developed for classical statistical systems which are
described by {\em positive} probability distribution functions of infinite
variables. FQH states are described by their ground state wave functions which
are {\em complex} functions of infinite variables. Thus it is not surprising
that FQH states contain addition structures (or a new kind of orders) that
cannot be described by broken symmetries and the Landau's theory. From this
point of view, we see that {\em any} quantum states may contain new kind of
orders that are beyond symmetry characterization. Such kind of orders was
studied in \Ref{Wqoslpub} and was called quantum order.  Since we cannot use
order parameter to describe quantum orders, a new mathematical object -
projective symmetry group (PSG) - was introduced\cite{Wqoslpub} to
characterize them. The topological order is a special case of quantum order -
a quantum order with a finite energy gap.

One may ask why do we need to introduce a new concept quantum order?  What use
can it have?  To answer such a question, we would like to ask why do we need
the concept of symmetry breaking? Is the symmetry breaking description of
classical order useful? Symmetry breaking is useful because (A) it leads to a
classification of crystal orders, and (B) it determines the structure of low
energy excitations without the needs to know the details of a
system.\cite{N6080,G6154}  The quantum order and its PSG description are
useful in same sense: (A) PSG can classify different quantum states that have
the same symmetry,\cite{Wqoslpub} and (B) quantum orders determine the
structure of low energy excitations without the needs to know the details of a
system.\cite{Wqoslpub,Wlight,WZqoind} The main difference between classical
orders and quantum orders is that classical orders produce and protect gapless
Nambu-Goldstone modes\cite{N6080,G6154} which is a bosonic excitation, while
quantum orders can produce and protect gapless gauge bosons and gapless
fermions.  Fermion excitations (with gauge charge) can even appear in pure
bosonic models as long as the boson ground state has a proper quantum
order.\cite{Wsrvb,K9721,Ktoap}

Those amazing properties of quantum orders could fundamentally change our
views on the universe and its elementary building blocks. The believed
``elementary'' particles, such as photons, electrons, \etc, may not be
elementary after all. Our vacuum may be a bosonic state with a non-trivial
quantum order where the ``elementary'' gauge bosons and the ``elementary''
fermions actually appear as the collective excitations above the quantum
ordered ground state. It may be the quantum order that protects the lightness
of those ``elementary'' particles whose masses are $10^{20}$ below the natural
mass - the Plank mass.  Those conjectures are not just wild guesses. A quantum
ordered state for a lattice spin model has been constructed\cite{Wlight} which
reproduces a complete QED with light, electrons, protons, atoms, ...

The concept of topological/quantum order is also useful in the field of
quantum computation.  People has been designing different kinds of quantum
entangled states to perform different computing tasks. When number of qubits
becomes larger and larger, it is more and more difficult to understand the
pattern of quantum entanglements.  One needs a theory to characterize
different quantum entanglements in many-qubit systems. The theory of
topological/quantum order\cite{Wtoprev,Wqoslpub} is just such a theory. In
fact topological/quantum orders can be viewed as patterns of quantum
entanglements and gauge bosons the fluctuations of quantum entanglements.
Also the robust topological degeneracy in topological ordered
states discovered in \Ref{Wtop} can be used in fault-tolerant quantum
computation.\cite{K9721}

It is hard to convince people about the usefulness of quantum orders when
their very existence is in doubt. However, a growing list of soluble or quasi
soluble models\cite{RS9173,K9721,MS0181,MSP0228,Ktoap,SM0220} indicates that
topological/quantum order do exist beyond FQH states. In particular, Kitaev
has constructed exactly soluble spin models that realize both topological
orders (\ie quantum orders with finite energy gap) and gapless quantum
orders.\cite{Ktoap}

In this paper, we study an exact soluble spin-1/2 model on square
lattice: $ H=16g \sum_{\v i} S^y_{\v i} S^x_{\v i+\hat{\v x}} S^y_{\v
i+\hat{\v x}+\hat{\v y}} S^x_{\v i+\hat{\v y}}$.  We find that the ground
states for $g<0$ and $g>0$ have the same symmetry but different quantum
orders.  The PSG's for those quantum ordered states are identified.  The phase
transition at $g=0$ represents a new kind of phase transitions that changes
quantum orders but not symmetry.  We show that the projective
construction that is used to construct quantum ordered ground
states\cite{BZA8773,BA8880,WWZcsp,Wsrvb,MF9400} not only gives us exact ground
states for our model, but also all the exact excited states. Through this
soluble model, we hope to put quantum order and its PSG description on a
firm ground. 

We would like to mention that the above spin-1/2 model, having one spin per unit
cell, is different from Kitaev's exact soluble spin-1/2 models on the
links of square lattice and on the sites of honeycomb lattice (which have two
spins per unit cell).\cite{K9721,Ktoap}  However, the $g<0$ version of the
above model corresponds to the low energy sector of Kitaev's honeycomb lattice
model in the $J_z >> J_x, J_y$ limit.\cite{Ktoap,K}

\noindent
{\bf Quantum orders in spin-1/2 and hard-core-boson models}

In this section, we are going to give a brief and general review of the PSG
description of quantum order. Readers who are interested in the exact
soluble model can go directly to the next section.
Let us consider a spin-1/2 system on a square lattice. 
Such a system can be viewed as a hard-core-boson model
if we identify $|\downarrow\>$ state as zero-boson state $|0\>$ and
$|\uparrow\>$ state as one-boson state $|1\>$.
In the follow we will use the boson picture to describe our model.

%For different boson Hamiltonians, our hard-core-boson system can have different
%ground state wave functions which may contain different quantum orders.  To
%study the possible quantum orders that may appear in our boson system, we
%follow the approach introduced in \Ref{Wqoslpub} and use the projective
%construction to obtain boson ground state wave functions with
%non-trivial quantum orders.

To construct quantum ordered (or entangled) many-boson wave functions, we
will use projective construction. We first introduce
a ``mean-field'' fermion Hamiltonian:\cite{Wqoslpub}
\begin{equation}
\label{Hmean}
 H_{mean}=
\sum_{\<\v i\v j\>} \left(
\psi^\dag_{I,\v i} \chi_{\v i\v j}^{IJ} \psi_{J,\v j}
+\psi^\dag_{I,\v i} \eta_{\v i\v j}^{IJ} \psi^\dag_{J,\v j}
+h.c. \right)
\end{equation}
where $I,J=1,2$.  We will use $\chi_{\v i\v j}$ and $\eta_{\v i\v j}$ to
denote the $2\times 2$ complex matrices whose elements are $\chi_{\v i\v
j}^{IJ}$ and $\eta_{\v i\v j}^{IJ}$.  
%We note that $\chi_{\v i\v j}$ and $\eta_{\v i\v j}$ satisfy
%$\chi^\dag_{\v i\v j} =\chi_{\v j\v i}$ and 
%$\eta_{\v j\v i} =-\eta_{\v i\v j}$.
Let $|\Psi^{(\chi_{\v i\v j}, \eta_{\v i\v j})}_{mean}\>$ be the ground state
of the above free fermion Hamiltonian, then a many-body boson wave function
can be obtained
\begin{equation}
\label{Phichieta}
 \Phi^{(\chi_{\v i\v j}, \eta_{\v i\v j})}(\v i_1, \v i_2 ...)=
\<0|\prod_{n}b(\v i_n)|\Psi^{(\chi_{\v i\v j},\eta_{\v i\v j})}_{mean}\>
\end{equation}
where
\begin{equation}
\label{bpsi}
 b(\v i) = \psi_{1,\v i}\psi_{2,\v i}
\end{equation}

According to \Ref{Wqoslpub}, the quantum order in the boson wave function
$\Phi^{(\chi_{\v i\v j}, \eta_{\v i\v j})}(\{\v i_n\})$ can be (partially)
characterized by projective symmetry group (PSG). To define PSG, we first
discuss two types of transformations. The first type is $SU(2)$ gauge
transformation
\begin{equation}
 (\psi_{\v i} , \chi_{\v i\v j}, \eta_{\v i\v j} )\to
(G(\v i)\psi_{\v i},G(\v i)\chi_{\v i\v j}G^\dag(\v j),
G(\v i)\eta_{\v i\v j}G^T(\v j))
\end{equation}
where $G(\v i)\in SU(2)$.
We note that the physical boson wave function $\Phi^{(\chi_{\v i\v
j}, \eta_{\v i\v j})}(\{\v i_n\})$ is invariant under the above $SU(2)$
gauge transformations.  The second type is the usual
symmetry transformation, such as the translations $T_x$: $\v i\to \v i -
\hat{\v x}$, $T_y$: $\v i\to \v i - \hat{\v y}$.
%, parities $P_x$: $(i_x,
%i_y)\to (-i_x, i_y)$, $P_y$: $(i_x, i_y)\to (i_x, -i_y)$.  
A generic transformation is a combination of the above two types, say
$GT_x(\chi_{\v i\v j}) = 
G(\v i)\chi_{\v i-\hat{\v x},\v j-\hat{\v x}}G^\dag(\v j)$.
The PSG for an ansatz $(\chi_{\v i\v j}, \eta_{\v i\v j})$ is formed by all
the transformations that leave the ansatz invariant.  

Every PSG contains a special
subgroup, which is called the invariant gauge group (IGG). An IGG is
formed by pure gauge transformations that leave the ansatz unchanged
%\begin{align}
$ IGG\equiv \{ G |\  
\chi_{\v i\v j}= G(\v i) \chi_{\v i\v j} G^\dag(\v j),
\eta_{\v i\v j}= G(\v i) \eta_{\v i\v j} G^T(\v j)
 \}$. 
%\end{align}
One can show that PSG, IGG, and the symmetry group (SG) of the many-boson wave
function are related: $PSG/IGG=SG$.\cite{Wqoslpub}

Different quantum orders in the ground states of our boson system are
characterized by different PSG's.  In the following we will concentrate on the
simplest kind of quantum orders whose PSG has a IGG=$Z_2$. We will call those
quantum states $Z_2$ quantum states.  We would like to ask how many different
$Z_2$ quantum states are there that have translation symmetry.  According to
our PSG characterization of quantum orders, the above physical question
becomes the following mathematical question: how many different PSG's are
there that satisfy $PSG/Z_2=$translation symmetry group. This problem has been
solved in \Ref{Wqoslpub}. The answer is 2 for 2D square lattice. Both PSG's
are generated by three elements $\{G_xT_x, G_yT_y, G_g\}$, where $G_g$ is a
pure gauge transformation that generates the $Z_2$ IGG: $IGG=\{1, G_g\}$.  The
gauge transformations in the three generators for the first $Z_2$ PSG are
given by
\begin{align}
\label{Z2APSG}
G_g(\v i) = -1,\ \ \
G_x(\v i) = 1, \ \ \ G_y(\v i) = 1.
\end{align}
Such a PSG will be called a Z2A PSG. The quantum states characterized by Z2A
PSG will be called Z2A quantum states.  
For the second $Z_2$ PSG, we have
\begin{align}
\label{Z2BPSG}
G_g(\v i) = -1, \ \ \
G_x(\v i) = 1, \ \ \  G_y(\v i) = (-1)^{i_x}.
\end{align}
Such a PSG will be called a Z2B PSG.
%The many-boson wave functions that contain the above two quantum orders are
%described by the following two ansatz:
%\begin{equation}
%\label{Z2A}
% \chi_{\v i,\v i+\v m} = \chi_{\v m}, \ \ \ 
% \eta_{\v i,\v i+\v m} = \eta_{\v m},
%\end{equation}
%and
%\begin{equation}
%\label{Z2B}
% \chi_{\v i,\v i+\v m} = (-1)^{m_x i_y} \chi_{\v m}, \ \ \
% \eta_{\v i,\v i+\v m} = (-1)^{m_x i_y} \eta_{\v m}.
%\end{equation}
%The above two ansatz lead to two explicit many-boson wave functions as
%indicated by \Eq{Phichieta}.  According to the theory of quantum order, the
%two  many-boson wave functions contain different quantum orders: Z2A and Z2B,
%and correspond to different quantum phases despite they have the same
%symmetry. One cannot continuously change one ground state to the other without
%encounter a phase transition.

%We would like to remark that the existence of only two projective extensions
%of the translation symmetry group by $Z_2$ (\ie the two PSG's discussed above)
%does not mean there are only two different quantum orders which have the
%translation symmetry and only the translation symmetry. As a partial and
%incomplete characterization, a PSG cannot resolve all the quantum orders. Two
%different quantum orders can share a same PSG.  However, the existence of two
%different PSG's does imply that there are at least two different quantum
%orders that are beyond the symmetry characterization.

If we increase the symmetry of the boson wave function, there can be more
different quantum orders.  A classification of quantum orders for spin-1/2
system on a square lattice is given in \Ref{Wqoslpub} where
hundreds of different quantum orders with the translation, parity, and
time-reversal symmetries were found.
%First we consider a SG generated by translations and parities
%$\{T_x,T_y,P_x,P_y\}$, where $P_x$: $(i_x,i_y)\to (-i_x,i_y)$ and $P_y$:
%$(i_x,i_y)\to (i_x,-i_y)$. Using the method developed in \Ref{WqoslHP}, we find
%there are 32 projective extensions of such a symmetry group by $Z_2$.  The 32
%PSG's are generated by five generators $\{G_xT_x, G_yT_y, G_{px}P_x,
%G_{py}P_y, G_g\}$. The gauge transformations in the five generators are given
%by \begin{align} G_x(\v i) =& \tau^0, & G_y(\v i) =& \tau^0, \nonumber\\
%G_{px}(\v i) =& \eta_{xpx}^{i_x} \eta_{ypx}^{i_y} \tau^0, & G_{py}(\v i) =&
%\eta_{xpy}^{i_x} \eta_{ypy}^{i_y} \tau^0 \nonumber\\ G_g(\v i) =& -\tau^0 ,
%\end{align} and \begin{align} G_x(\v i) =& \tau^0, & G_y(\v i) =& (-1)^{i_x}
%\tau^0, \nonumber\\ G_{px}(\v i) =& \eta_{xpx}^{i_x} \eta_{ypx}^{i_y} \tau^0,
%& G_{py}(\v i) =& \eta_{xpy}^{i_x} \eta_{ypy}^{i_y} \tau^0 \nonumber\\ G_g(\v
%i) =& -\tau^0 , \end{align} where $\eta$'s can independently take one of two
%possible values $\pm 1$.  The 32 different PSG's suggest that 32 possible
%$Z_2$ quantum states that have the translation and the parity symmetries.

In the following we will study an exact soluble model whose ground states
realize some of the constructed quantum orders.  Our model has the following
nice property: the projective construction \Eq{Phichieta} gives us all the
energy eigenstates for some proper choices of $\chi_{\v i\v j}$ and $\eta_{\v
i\v j}$. All the energy eigenvalues can be calculated exactly.

\noindent
{\bf Exact soluble models on 2D square lattice}

Our construction is motivated by
Kitaev's construction of soluble spin-1/2
models on honeycomb lattice.\cite{Ktoap}  The key step in both constructions
is to find a system of commuting operators. Let
$ \hat U^a_{\v i\v j} \equiv \la^T_{\v i} U^a_{\v i\v j} \la_{\v j}$,
where $\v i, \v j$ label lattice sites, $a$ is an integer index,
$U^a_{\v i\v j}$ is an $n\times n$ matrix satisfying 
%\begin{equation}
%(U^a_{\v i\v j})^T= - U^a_{\v j\v i} 
%\end{equation}
$(U^a_{\v i\v j})^T= - U^a_{\v j\v i}$, 
and $\la_{\v i}^T=( \la_{1,\v i} , \la_{2,\v i} ,..., \la_{n,\v i} )$ 
is a $n$-component Majorana fermion operator satisfying
$ \{ \la_{a,\v i},\la_{b,\v j} \} = 2\del_{ab}\del_{\v i\v j}$.
We require that all $\hat U^a_{\v i\v j}$ to commute with each other:
$ [ \hat U^a_{\v i_1\v i_2}, \hat U^b_{\v j_1\v j_2} ] = 0$,
which can be satisfied iff
% \begin{align}
% U_{ab}V_{cd} \nonumber\\
%  [ab,cd] = [a,cd]b+a[b,cd]
% = -c{a,d}b+{a,c}db -ac{b,d}+a{b,c}d
% = -cb \del_{ad} +db\del_{ac} - ac\del_{bd} + ad\del_{bc}  \nonumber\\
%   - U_{cb}V_{ac} + U_{cb}V_{ca} - U_{ac}V_{bc} + V_{ac}U_{cb} 
% = - V_{ac}U_{cb} + V_{ca}U_{cb} - U_{ac}V_{bc} + U_{ac}V_{cb} 
% = - V_{ac}U_{cb} - V_{ac}U_{cb} + U_{ac}V_{cb} + U_{ac}V_{cb} 
% = - VU - VU + UV + UV = 2[U,V]  \nonumber\\
% (U^{i1i2}_{a,c}\del_{i1 i}\del_{i2 k}-U^{i1i2}_{c,a}\del_{i1 k}\del_{i2 i}) 
% (U^{j1j2}_{c,b}\del_{j1 k}\del_{j2 j}-U^{j1j2}_{b,c}\del_{j1 j}\del_{j2 k}) 
% =  U^{i1i2} U^{j1j2}\del_{i1 i}\del_{j2 j}\del_{i2 j1}
%  - U^{i1i2} U^{j1j2T}\del_{i1 i}\del_{j1 j}\del_{i2 j2}
%  - U^{i1i2T} U^{j1j2}\del_{i2 i}\del_{j2 j}\del_{i1 j1}
%  + U^{i1i2T} U^{j1j2T}\del_{i2 i}\del_{j1 j}\del_{i1 j2}  \nonumber\\
% U^{i1i2} V^{i2i3}
% => U^{i1i2} V^{i2i3}\del_{i1 i}\del_{i3 j}
% V^{i2i3} U^{i1i2} 
% => V^{i2i3T} U^{i1i2T}\del_{i3 i}\del_{i1 j}
% U^{i1i2} V^{i2i1}
% =>  U^{i1i2} V^{i2i1}\del_{i1 i}\del_{i1 j}
%   + U^{i1i2T} V^{i2i1T}\del_{i2 i}\del_{i2 j}
% V^{i2i1} U^{i1i2}
% =>  V^{i2i1} U^{i1i2}\del_{i2 i}\del_{i2 j}
%   + V^{i2i1T} U^{i1i2T}\del_{i1 i}\del_{i1 j}
% U^{i1i1} V^{i1i2}
% => U^{i1i1} V^{i1i2}\del_{i1 i}\del_{i2 j}
%   -U^{i1i1T} V^{i1i2}\del_{i1 i}\del_{i2 j}
% V^{i1i2} U^{i1i1}
% => V^{i1i2T} U^{i1i1T}\del_{i2 i}\del_{i1 j}
%   -V^{i1i2T} U^{i1i1}\del_{i2 i}\del_{i1 j}
% \end{align}
\begin{align}
\label{UU0}
U^a_{\v i_1\v i_2} U^b_{\v i_2\v i_3} = &0 , \ \ \
U^a_{\v i_1\v i_2} U^b_{\v i_2\v i_1}
= (U^a_{\v i_1\v i_2} U^b_{\v i_2\v i_1})^T  \nonumber\\
U^a_{\v i_1\v i_1} U^b_{\v i_1\v i_2} = &0  
\end{align}
where $\v i_1$, $\v i_2$, $\v i_3$ are all different.

Let $\{|1\>, ..., |4\>\} $ be a basis of a $4$ dimensional real linear space.
Then the following $U_{\v i\v j}$ on a square lattice 
$U_{\v i,\v i+\hat{\v x}} =   |1\>\<3| $, 
$U_{\v i,\v i-\hat{\v x}} =  -|3\>\<1| $, 
$U_{\v i,\v i+\hat{\v y}} =   |2\>\<4| $, and 
$U_{\v i,\v i-\hat{\v y}} =  -|4\>\<2| $
form a solution of \Eq{UU0}. We find that 
\begin{equation} 
\hat U_{\v i,\v i+\hat{\v x}}=\la_{1,\v i}\la_{3,\v i+\hat{\v x}}, \ \ 
\hat U_{\v i,\v i+\hat{\v y}}=\la_{2,\v i}\la_{4,\v i+\hat{\v y}} 
\end{equation} 
form a commuting set of operators.

After obtaining a commuting set of operators, we can easily see that
the following Hamiltonian
\begin{align}
\label{H}
 H = g\sum_{\v i} \hat F_{\v i}, \ \ \
\hat F_{\v i} = 
\hat U_{\v i,\v i_1} 
\hat U_{\v i_1, \v i_2} 
\hat U_{\v i_2,\v i_3} 
\hat U_{\v i_3,\v i} 
\end{align}
commutes with all the $\hat U_{\v i\v j}$'s,
where 
$\v i_1 = \v i+\hat{\v x}$,
$\v i_2 = \v i+\hat{\v x}+\hat{\v y}$, and $\v i_3 = \v i+\hat{\v y}$.  We will call 
$\hat F_{\v i}$ a $Z_2$
flux operator.  Let $|s_{\v i\v j}\>$  be the common eigenstate of $\hat U_{\v
i\v j}$ with eigenvalue $s_{\v i\v j}$.  Since $(\hat U_{\v i\v j})^2=-1$,
$s_{\v i\v j}$ satisfies $s_{\v i\v j}=\pm i$ and $s_{\v i\v j}=-s_{\v j\v i}$.
$|s_{\v i\v j}\>$ is also an energy eigenstate with energy
\begin{align} 
\label{Eng}
E = g\sum_{\v i} F_{\v i}
,\ \ \ 
F_{\v i} = s_{\v i,\v i_1} s_{\v i_1, \v i_2} s_{\v i_2,\v i_3} s_{\v i_3,\v i} 
\end{align} 

Let us discuss the Hilbert space within which
the above $H$ acts.  
On each site, we group $\la_{1,2,3,4}$
into two fermion operators
\begin{equation}
 \psi_{1,\v i} = \la_{1,\v i} +i\la_{3,\v i}, \ \ \
 \psi_{2,\v i} = \la_{2,\v i} +i\la_{4,\v i}
\end{equation}
$\psi_{1,2}$ generate a four dimensional Hilbert space on each site.  Let us
assume the 2D square lattice to have $N_s$ lattice sites and a periodic
boundary condition in both directions.  Since there are total of $2^{2N_s}$
different choices of $s_{\v i\v j}$ (two choices for each link), the states
$|s_{\v i\v j}\>$ exhaust all the $4^{N_s}$ states in the Hilbert space.  Thus
the common eigenstates of $\hat U_{\v i\v j}$ is not degenerate and the above
approach allows us to obtain all the eigenstates and eigenvalues of the $H$.
%The explicit form of $|s_{\v i\v j}\>$ can also be found \begin{equation}
%|s_{\v i\v j}\>  =\prod_{\<\v i\v j\>} (1-s_{\v i\v j}\hat U_{\v i\v j})|0\>
%\end{equation} where $\<\v i\v j\>$ represents nearest-neighbor links and
%$|0\>$ is the state with no fermions.  The above is wrong since $\prod_{\<\v
%i\v j\>} (1-s_{\v i\v j}\hat U_{\v i\v j})|0\>$ can be zero for some $s_{\v
%i\v j}$

We note that the Hamiltonian $H$ can only change the fermion number on each
site by an even number. Thus the $H$ acts within a subspace which has an even
number of fermions on each site. The subspace has only two states per site.
When defined on the subspace, $H$ actually describes a spin-1/2 or a hard-core
boson system under the operator mapping \Eq{bpsi}.  The subspace is formed by
states that are invariant under local $Z_2$ gauge transformations: $\psi_{I\v
i} \to G(\v i)\psi_{I\v i}$, $G(\v i)=\pm 1$.  We will call those states
physical states and call the subspace the physical Hilbert space.

Since $\hat U_{\v i\v j}$ do not act within the physical Hilbert space, they
do not have definite values for physical states. However, we note that the
$Z_2$ flux operator $\hat F_{\v i}$ act within the physical Hilbert space. The
$Z_2$ flux operators commute with each other and the $H$ is a function of
$\hat F_{\v i}$.  To obtain the common eigenstates of the $Z_2$ flux operators
in the physical Hilbert space, we note that the $Z_2$ flux operators are
invariant under the $Z_2$ gauge transformation: $\psi_{I\v i} \to G(\v
i)\psi_{I\v i}$.  A $Z_2$ gauge transformation changes one eigenstate $|s_{\v
i\v j}\>$ to another eigenstate $(\prod_{\v i} G_{\v i}^{\hat N_{\v i}})
|s_{\v i\v j}\>\propto |G(\v i)s_{\v i\v j}G(\v j)\>$, where $\hat N_{\v i}$
is the fermion number operator at site $\v i$.  We will call those two
eigenstates gauge equivalent.  The common eigenstates of the $Z_2$ flux
operators within the physical Hilbert space can be obtained by summing over all
the gauge equivalent eigenstates with equal amplitude:
\begin{equation}
\label{phyG}
|s_{\v i\v j}\>_{phy} \equiv \sum_G 
(\prod_{\v i} G_{\v i}^{\hat N_{\v i}}) |s_{\v i\v j}\>
\end{equation}
(Note $|s_{\v i\v j}\>_{phy}$ can also be viewed as the projection of
$|s_{\v i\v j}\>$ onto the physical Hilbert space.)

Let us count the physical states that can be constructed this way, again
assuming a periodic boundary condition in both directions.  We note that the
terms in the above sum can be grouped into pairs, where the $Z_2$ gauge
transformations in a pair differs by a uniform $Z_2$ gauge transformation
$G(\v i)=-1$. We see that the sum of the two terms in a pair is zero if the
total number of the fermions $N_f = \sum N_{\v i}$ is odd.  Thus only states
$|s_{\v i\v j}\>$ with even number of fermions leads to physical states in the
above construction.  For states with even number of fermions, there are only
$2^{N_s}/2$ distinct terms in the above sum. Thus each physical eigenstate
comes from $2^{N_s}/2$ gauge equivalent eigenstates. Since there are
$4^{N_s}/2$ states with even number of fermions, we find the number of
physical states is $2^{N_s}$, which is the dimension of the physical Hilbert
space.  Thus we can obtain all the eigenstates and eigenvalues of the $H$ in
the physical Hilbert space from our construction.

Let us introduce a notion of $Z_2$ flux configuration. Among all the possible
$s_{\v i\v j}$'s, we can use the $Z_2$ gauge transformation $s_{\v i\v j}\to
G(\v i)s_{\v i\v j}G(\v j)$ to define an equivalence relation.  A $Z_2$ flux
configuration is then an equivalence class under the $Z_2$ gauge transformation.
From the above picture, we see that for every choice of $Z_2$ flux
configuration $F_{\v i} = \pm 1$, we can choose a $s_{\v i\v j}$ that
reproduces the $Z_2$ flux on each plaquette.  If the state $|s_{\v i\v j}\>$
has even numbers of fermions, it will leads to a physical eigenstate (see
\Eq{phyG}).  The energy of the physical eigenstate is given by \Eq{Eng}.  The
explicit many-boson wave function of the eigenstate is given by $ \Phi(\{\v
i_n\})= \<0|\prod_{n}b(\v i_n)|\Psi_{mean}\> $ where
$|\Psi_{mean}\>$ is the ground state of
\begin{equation}
\label{Hmean1}
H_{mean} = \sum_{\<\v i\v j\>} \left( s_{\v i\v j}\hat U_{\v i\v j}+h.c.\right) 
\end{equation}
Here we see that all the eigenstates of our model can be obtained from the
projective construction \Eq{Phichieta}.  
%(We would like to remark that some
%$Z_2$ flux configurations correspond to states with odd numbers of fermions.
%Those $Z_2$ flux configurations do not correspond to any physical states.)

\noindent
{\bf Physical properties}

% Calculate Hamiltonian in term of spin operators:
% F = la1la3*la2la4*la3la1*la4la2 
%  = -(la2la1)(la3la2)(la4la3)(la1la4)
%  Case A:
% psi1 = la1+ila3   psi2 = la2+ila4
% la2la1 = (psi2+psi2d)(psi1+psi1d)
%        = psi2psi1 + psi2dpsi1d + ...
% la3la2 = -i(psi1-psi1d)(psi2+psi2d)
%        = -i psi1psi2 + i psi1dpsi2d + ...
% la4la3 = -(psi2-psi2d)(psi1-psi1d)
%        = - psi2psi1 - psi2dpsi1d + ...
% la1la4 = -i(psi1+psi1d)(psi2-psi2d)
%        = -i psi1psi2 + i psi1dpsi2d + ...
%  Case B:
% psi1 = la1+ila2   psi2 = la3+ila4
% la2la1 = -i(psi1-psi1d)(psi1+psi1d)
%        = -i psi1psi1d + i psi1dpsi1
% la3la2 = -i(psi2+psi2d)(psi1-psi1d)
%        = -i psi2psi1 + i psi2dpsi1d + i psi2psi1d - i psi2dpsi1
% la4la3 = -i(psi2-psi2d)(psi2+psi2d)
%        = -i psi2psi2d + i psi2dpsi2
% la1la4 = -i(psi1+psi1d)(psi2-psi2d)
%        = -i psi1psi2 + i psi1dpsi2d + ...

In terms of the hard-core-boson operator \Eq{bpsi}, the Hamiltonian \Eq{H}
of our soluble model has a form
$ H=-g \sum_{\v i}
(b-b^\dag)_{\v i}
(b+b^\dag)_{\v i+\hat{\v x}}
(b-b^\dag)_{\v i+\hat{\v x}+\hat{\v y}}
(b+b^\dag)_{\v i+\hat{\v y}} $.
In terms of spin-1/2 operator $\tau^x=b+b^\dag$ and $\tau^y=i(b-b^\dag)$, the
Hamiltonian has a form
\begin{equation}
 H=g \sum_{\v i} \hat F_{\v i}
 =g \sum_{\v i} \tau^y_{\v i} \tau^x_{\v i+\hat{\v x}}
\tau^y_{\v i+\hat{\v x}+\hat{\v y}} \tau^x_{\v i+\hat{\v y}}
\end{equation}
In the bulk,  the above model corresponds to the following simple Ising
model $H_{Ising}=g \sum_{\v i} \tau_{\v i}^z$.  However, the two models are
different for finite systems and for systems with edges (see discussions below).

When $g<0$, the ground state of our model is given by $Z_2$ flux configuration
$F_{\v i} = 1$. To produce such a flux, we can choose $s_{\v i,\v i+\hat{\v
x}} = s_{\v i,\v i+\hat{\v y}} = i$. In this case, \Eq{Hmean1} becomes
\Eq{Hmean} with $-\eta_{\v i,\v i+\hat{\v x}} = \chi_{\v i,\v i+\hat{\v x}}
=1+\tau^z$ and $ -\eta_{\v i,\v i+\hat{\v y}} = \chi_{\v i,\v i+\hat{\v y}} =
1-\tau^z $.  The PSG for the above ansatz turns out to be the Z2A PSG in
\Eq{Z2APSG}.  Thus the ground state for $g<0$ is a Z2A state.  

When $g>0$, the ground state is given by configuration $F_{\v i} = -1$ which
can be produced by $(-)^{i_y}s_{\v i,\v i+\hat{\v x}} = s_{\v i,\v i+\hat{\v
y}} = i$.  The ansatz now has a form $-\eta_{\v i,\v i+\hat{\v x}} = \chi_{\v
i,\v i+\hat{\v x}} =(-)^{i_y}(1+\tau^z)$ and $ -\eta_{\v i,\v i+\hat{\v y}} =
\chi_{\v i,\v i+\hat{\v y}} = 1-\tau^z$.  Its PSG is the Z2B PSG in
\Eq{Z2BPSG}.  Thus the ground state for $g>0$ is a Z2B state.  

Both the Z2A and the Z2B states has translation $T_{x,y}$, parity $P_{xy}:
(i_x,i_y)\to(i_y,i_x)$, and time reversal (since $\chi_{\v i\v j}$ and
$\eta_{\v i\v j}$ are real) symmetries.  The low energy excitations in both
states are $Z_2$ vortices generated by flipping the signs of an even numbers
of $F_{\v i}$'s.  Those $Z_2$ vortex excitations have a finite energy gap
$\Del =2|g|$.
% and behave like bosons. 
The low energy sector of our model is identical to the low energy sector of a
$Z_2$ lattice gauge theory.  However, our model is not equivalent to a $Z_2$
lattice gauge theory. This is because a $Z_2$ lattice gauge theory has
$4\times 2^{N_s}/2$ states on a torus while our model has $2^{N_s}$ states.

Due to the low energy $Z_2$ gauge structure, both the Z2A and the Z2B states
have four degenerate ground states on an even by even lattice with periodic
boundary condition. The degeneracy is topological and is protected by the
topological order in the two states. The degeneracy is robust against arbitrary
perturbations. This is because perturbations are $Z_2$ gauge invariant
physical operators which cannot break the low energy $Z_2$ gauge structure
that produces the degeneracy.\cite{Wsrvb,Wtop}

On an even by even lattice, the ground state energy per site is given by
$-|g|$. The singularity at $g=0$ implies a phase transition between two states
with the same symmetry and the same ground state degeneracy!  Thus the
transition is a new type of continuous transitions that only changes the
quantum orders (from Z2A to Z2B).  The transition is continuous since gap
vanishes at $g=0$.  Although the Z2A and the Z2B states share many common
properties on an even by even lattice, the two states are quite different on
an odd by odd lattice. On an odd by odd lattice, the Z2A state has an energy
$-|g|N_s$ and a $2$ fold degeneracy, while the Z2B state, containing a single
$Z_2$ vortex to satisfy the constraint $\prod_{\v i} F_{\v i}=1$, has an
energy $-|g|(N_s-2)$ and a $2N_s$ fold degeneracy.  On lattice with edges in
$(1,0)$ and/or $(0,1)$ directions, our model has $\sim 2^{N_{edge}/2}$ gapless
edge states, where $N_{edge}$ is the number of edge sites.

%\begin{acknowledgments}
This research is supported by NSF Grant No. DMR--01--23156
and by NSF-MRSEC Grant No. DMR--98--08941.
%\end{acknowledgments}

%\bibliography{/home/wen/bib/wencross,/home/wen/bib/htc,/home/wen/bib/misc,/home/wen/bib/fqh,/home/wen/bib/part,/home/wen/bib/qcom,/home/wen/bib/publst} 

\begin{thebibliography}{23}
\bibitem[{\citenamefont{Kitaev}(2002)}]{K}
I would like thank Kitaev for pointing this out to me.
\end{thebibliography}

\end{document}